\documentclass[12pt]{iopart}
\usepackage{iopams}
\usepackage{graphicx}
\usepackage{subcaption}
\usepackage{color}
\usepackage{cite}
\usepackage{hyperref}

\begin{document}

\title{How Soft is Too Soft? Tuning Order and Disorder in Dimeric Core-Soft Colloids with Bond Flexibility}

\author{Leandro B. Krott$^1$, Davi Felipe Kray Silva$^2$, A. de J. Ríos-Roldán$^3$, Victor M. Trejos$^4$, J. Antonio Moreno-Razo$^4$, Jos\'e Rafael Bordin$^2$}

\address{$^1$ Centro de Ciências, Tecnologias e Saúde, Campus Araranguá, Universidade Federal de Santa Catarina, Rua Pedro João Pereira, 150, CEP 88905-120, Araranguá, SC, Brasil.}
\address{$^2$ Departamento de F\'isica, Instituto de F\'isica e Matem\'atica, Universidade Federal de Pelotas, Caixa Postal 354, 96001-970, Pelotas-RS, Brasil.}
\address{$^3$ Departamento de Física, Universidad Autónoma Metropolitana-Iztapalapa, Mexico City, Mexico.}
\address{$^4$ Departamento de Química, Universidad Autónoma Metropolitana-Iztapalapa, Av. San Rafael Atlixco 186, Col. Vicentina, 09340 Ciudad de México, Mexico.}

\ead{jrbordin@ufpel.edu.br}

\begin{abstract}
We employ molecular dynamics simulations to explore how internal flexibility affects phase transitions in soft-matter systems composed of dimers interacting via a core-softened potential with two characteristic length scales. Monomers are connected by harmonic springs with varying stiffness, allowing us to tune the dimer rigidity from highly flexible to nearly rigid. Flexible dimers reproduce the behavior of monomeric systems, displaying well-defined BCC and HCP crystalline phases separated by a narrow amorphous region. As the bond stiffness increases, this amorphous phase gives way to a coexistence region between BCC and HCP structures. In the rigid limit, amorphous regions reemerge and expand, and high-density systems fail to crystallize 
completely,
instead forming mixed phases with HCP-like and disordered local environments. This transition arises from geometric frustration: rigid dimers are unable to adjust their internal configuration to optimize local packing, thereby suppressing crystallization and promoting amorphization. Our 
findings  reveal that bond flexibility is a key control parameter governing structural organization in core-softened colloidal and molecular systems, offering insights for the design of tunable soft materials.
\end{abstract}

\submitto{Journal of Physics: Condensed Matter}

\section{Introduction}

Understanding how internal structural features influence phase transitions is a key challenge in the statistical physics of complex fluids. In systems such as water~\cite{gallo21,gallo16,coronas2024}, colloidal suspensions~\cite{Hueckel2021,SolanoCabrera2025,Li2016}, and biological media~\cite{deGennes1992,Lee_2019,Pappu2023,Alexandrov2020}, phase behavior is governed not only by intermolecular interactions but also by internal degrees of freedom—such as shape anisotropy, bond flexibility, and rotational constraints. These internal properties often give rise to emergent phenomena, including structural frustration, amorphous ordering, and thermodynamic anomalies, which are not easily captured by classical models.

To study such complex systems, coarse-grained models offer a powerful approach by capturing essential physical behavior while greatly reducing the computational cost associated with fully atomistic simulations~\cite{yan06}. Among these, core-softened (CS) potentials—characterized by a repulsive core followed by a softened or attractive shoulder—have been extensively used to reproduce waterlike anomalies, including density maxima, diffusion anomalies, and structural transitions~\cite{hemmer70,jagla1999b,Scala2000,Skibinsky04}.

Beyond mimicking the behavior of water, CS potentials have been successfully employed to model a wide class of 
soft-matter systems that exhibit deformable surfaces and multi-scale interactions, 
offering a richer description than the classical hard-sphere paradigm~\cite{Royall2024}.
These systems include microgels, dendrimers, micelles, and polymer-grafted nanoparticles, where steric stabilization, osmotic effects, or compressed polymer brushes generate a short-range repulsion and a soft intermediate shoulder~\cite{Likos01,Louis2000,Trejos2018,Trejos2019}. Such features are central to determining their thermodynamic and structural behavior~\cite{Scotti2022,Vlassopoulos2014}.

Experimental studies have demonstrated that the effective interactions in systems of pure or grafted polyethylene glycol (PEG) colloids can be accurately described by core-softened potentials~\cite{colloid1,colloid2,Haddadi20}. These interactions typically feature a soft repulsive core followed by a region of reduced repulsion or weak attraction, often arising from entropic contributions or polymer-mediated forces. Computational studies have reported similar findings for polymer-grafted nanoparticles~\cite{marques2020a,Lafitte2014} and star polymers~\cite{Bos19}, where the overlap of polymer coronas naturally gives rise to effective potentials with two characteristic length scales. Core-softened models have proven remarkably
versatile in capturing a variety of phenomena observed in soft colloidal systems, including equilibrium clustering~\cite{Mladek2006}, reentrant melting~\cite{Likos2006,Bordin2023a}, and the formation of complex mesophases~\cite{Padilla2021,Kryuchkov2018,BordinSM2023}. Moreover, the tunability of the soft shoulder’s range and depth allows~\cite{TorresCarbajal2020,Trejos2020}
the parametrization of effective potentials based on experimental systems~\cite{Yethiraj2003}, providing predictive power for the design of new soft materials~\cite{Bassani2024}.

Most of these studies consider colloidal particles as spherically symmetric and rigid. A natural extension of these models involves introducing geometrical constraints by connecting monomers into dimers. Dimeric systems introduce additional internal degrees of freedom -- such as bond-length fluctuations and orientational anisotropy -- that couple to local packing constraints and can generate complex phase behavior~\cite{Glotzer2007,Zhou2024}. Previous studies on rigid dimers with core-softened interactions have reported extended solid regions, cluster formation, liquid-crystalline phases, and confinement-induced structural transitions~\cite{Barros_de_Oliveira_2010,Bordin2016,Bordin16c,Bordin17,KGB2016,Munao2015}. In two dimensions, core-softened dumbbells can also form complex ordered patterns absent in monomeric systems~\cite{Yang2019,Nogueira22}. While some studies have considered varying anisotropy~\cite{gava2014,Netz11,Munao2015b,Nogueira2023}, they all assume fixed internal distances and neglect bond flexibility—despite its relevance in real molecular systems such as polymers, amphiphiles, and biomacromolecules.

Despite its importance, the role of bond flexibility in the phase behavior of core-softened dimers remains largely unexplored. This omission is significant, as internal flexibility can drastically affect local packing, structural transitions, and crystallization efficiency. Furthermore, geometric frustration—arising when preferred local order is incompatible with long-range crystalline symmetry—can lead to the stabilization of amorphous solids, phase coexistence, or even complete suppression of crystallization~\cite{Simonov2020}. While rigid bonds amplify such frustration, flexible ones can relieve it by enabling local structural rearrangements~\cite{Palberg2016,Janicki2022}. As a result, dimer flexibility emerges as a critical parameter to tune the interplay between order and disorder in soft-matter systems.

In this work, we investigate how the rigidity of the dimer bond affects the thermodynamic and structural behavior of a fluid composed of dimers interacting via a core-softened potential with two characteristic length scales. By systematically varying the spring constant $k$ of the harmonic bond connecting the monomers, we interpolate between highly flexible and nearly rigid dimers. Using molecular dynamics simulations, we construct pressure–temperature phase diagrams, identify both first- and second-order phase transitions, and characterize the resulting solid phases using structural and thermodynamic observables. Our results reveal a rich diversity of structural regimes: flexible dimers exhibit well-defined crystalline arrangements similar to those found inmonomeric systems, while increasingly rigid dimers lead to extended amorphous regions, coexistence between HCP and disordered solids, and suppression of crystallization due to geometric frustration. These findings provides new insights into how internal degrees of freedom modulate phase behavior in anomalous fluids and offers a minimal yet robust framework for exploring frustration-driven glass formation, solid–solid transitions, and structural competition in soft-matter and molecular systems.

The paper is organized as follows: Sec.~\ref{model} introduces the model and simulation details. In Sec.~\ref{results}, we present and discuss the main results. Finally, Se.~\ref{conclusions} offers concluding remarks and perspectives for future work.

\section{The Model and Computational Details}\label{model}

The system consists of dumbbell-like particles composed of two CS particles connected by a harmonic bond with spring constant $k$,
\begin{equation}
U_k(r) = \frac{1}{2}k(r - \lambda)^2,
\end{equation}
allowing oscillations around the equilibrium bond length $\lambda = 1.0\sigma$. We considered several values of $k$—0.1, 1.0, 2.0, 3.5, 5.0, 10.0, 100.0, and 1000.0—to capture the entire range from flexible to rigid dimer behavior.

\begin{figure}[htp!]
    \centering
    \includegraphics[width=0.6\textwidth]{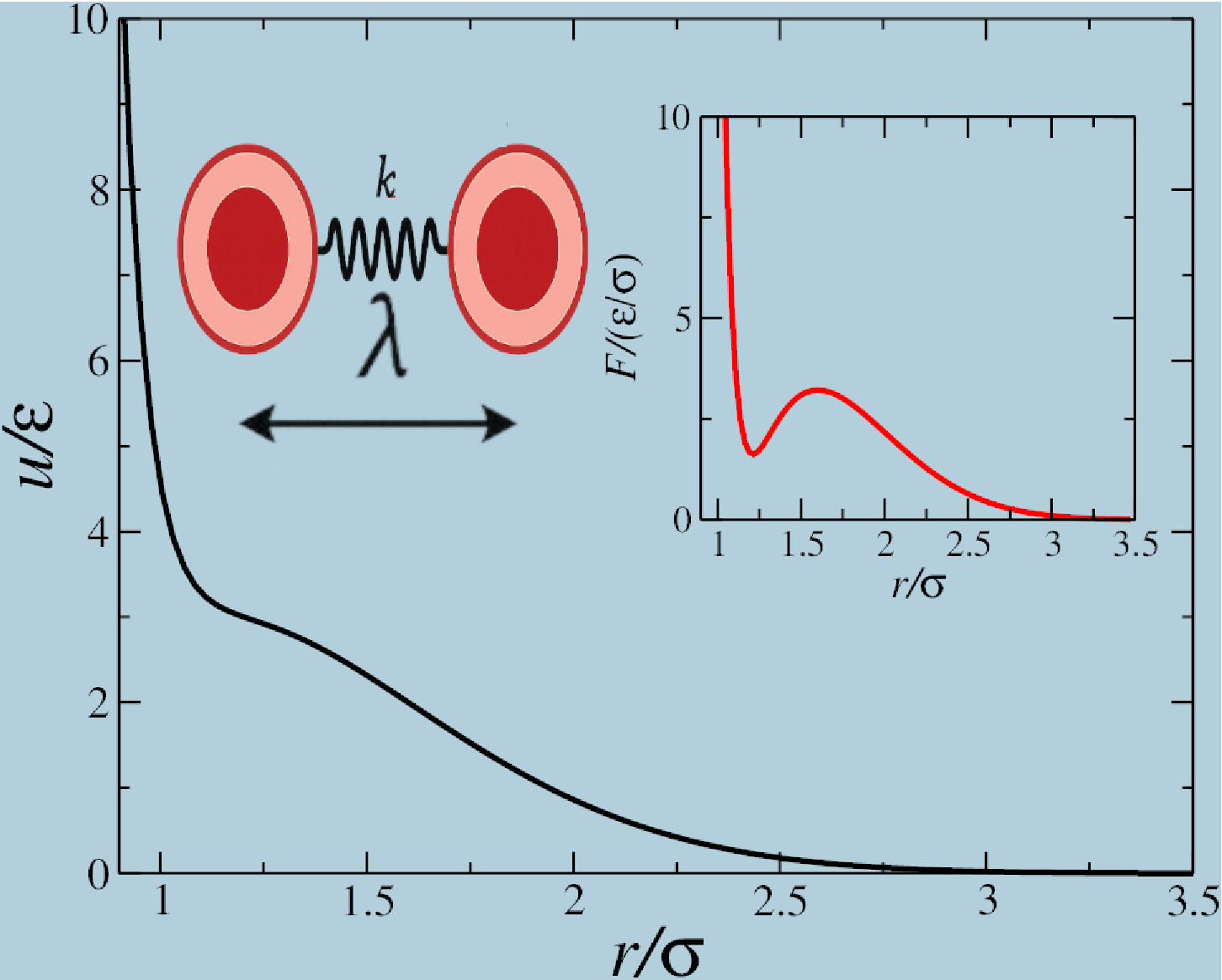} 
    \caption{Interaction potential as a function of the interparticle distance. The inset shows the force profile, $F = -\partial u/\partial r$. The molecular inset schematically represents the core-softened dimer, with monomers separated by a distance $\lambda$.}
    \label{potential}
\end{figure}

The interaction between non-bonded monomers is modeled using a core-softened potential with two characteristic length scales. This potential consists of a Lennard-Jones (LJ) term combined with a Gaussian well, defined as~\cite{BarrosDeOliveira2006,DeOliveira2006,Bordin2023a}:

\begin{equation}
u(r) = 4\varepsilon\left[\left(\frac{\sigma}{r}\right)^{12} - \left(\frac{\sigma}{r}\right)^6\right] + u_0 \exp\left[-\frac{1}{c_0^2}\left(\frac{r - r_0}{\sigma}\right)^2\right],
\label{eq:CS}
\end{equation}

\noindent where $r_0/\sigma = 0.7$, $u_0 = 5\varepsilon$, and $c_0^2 = 1.0$. Figure~\ref{potential} shows the shape of the core-softened potential, with the inset displaying the corresponding force, $F = -\partial u/\partial r$, as a function of the interparticle distance. The two characteristic length scales, located at approximately $r \approx 1.2$ and $r \approx 2.0$, are responsible for waterlike anomalies in the system. A cutoff radius of $r_c = 5.0$ was used in all simulations.

Molecular dynamics simulations in the canonical ensemble ($NVT$)~\cite{Frenkel} were performed using the LAMMPS package~\cite{lammps_citation}. The system consisted of 1000 dumbbells. Temperature was controlled using a Nosé–Hoover thermostat~\cite{hoover} with a damping parameter of $Q_T = 0.1$. The timestep was set to $\delta t = 0.001$, and the velocity-Verlet algorithm~\cite{verlet} was employed to integrate the equations of motion.

The system was initialized with random positions and velocities at the highest simulated temperature for each density. For each temperature step during cooling, the final configuration of the previous run was used as the starting point, ensuring quasi-static equilibration. Each state was equilibrated over $10^6$ timesteps, followed by data collection over another $10^6$ timesteps. The heating process followed a similar protocol, beginning from the final configuration at the lowest temperature obtained during cooling. System stability was verified by monitoring energy and pressure as functions of time.

To identify phase transitions, we analyzed both thermodynamic, dynamic and structural properties. The specific heat at constant volume, $C_V$, is a key thermodynamic quantity and is given by

\begin{equation}
C_V = \left( \frac{\partial U}{\partial T} \right)_V,
\end{equation}

\noindent where $U$ is the total potential energy. Discontinuities or sharp peaks in $C_V(T)$ indicate first-order transitions, while broader peaks suggest second-order or continuous transitions.

Structurally, we computed the radial distribution function $g(r)$ for the dimer center of mass and used it to calculate the translational order parameter $\tau$~\cite{Shell02,Errington01,Errington03}:

\begin{equation}
\tau = \int \left| g(\epsilon) - 1 \right| d\epsilon,
\end{equation}

\noindent where $\epsilon = r\rho^{1/3}$ and $\rho$ is the number density. $\tau$ ranges from zero (ideal gas) to higher values for ordered phases, making it a useful indicator of structural changes and transitions~\cite{DeOliveira2006,Krott13,Krott2015,Leo17,Bordin2018,Cardoso21,Bordin2023a}.

We also computed the excess entropy $s_{ex}$, whose leading contribution is the two-body term~\cite{raveche1971,Baranyai1989,Baranyai89}:

\begin{equation}
s_{2} = -2\pi \rho \int_{0}^{\infty} \left[g(r)\ln g(r) - g(r) + 1\right] r^2 dr,
\label{s2}
\end{equation}

\noindent and its cumulative form~\cite{Krek08,Cardoso21}:

\begin{equation}
C_{s_2}(r) = -2\pi\rho \int_0^r \left[g(r')\ln g(r') - g(r') + 1\right] r'^2 dr',
\label{cs2}
\end{equation}

\noindent which provides insights into longer-range ordering. Fluids exhibit converging $C_{s_2}(r)$ profiles, while solids show divergent behavior.

To investigate dynamic properties, we computed the diffusion coefficient $D$ using the Einstein relation:

\begin{equation}
D = \lim_{t \rightarrow \infty} \frac{\langle \Delta \vec{r}(t)^2 \rangle}{6t},
\end{equation}

\noindent where $\langle \Delta \vec{r}(t)^2 \rangle$ is the mean square displacement of the dimer's center of mass.

To distinguish between solid phases, we evaluated two local order parameters: the orientational order parameter (OOP) and the centrosymmetry parameter (CSP). The OOP for particle $i$ with $N_b$ neighbors is given by~\cite{steinhardt1983,Er01,DeOliveira2006,Yan05}:

\begin{equation}
q_l(i) = \sqrt{\frac{4\pi}{2l + 1} \sum_{m = -l}^{l} |q_{lm}|^2},
\end{equation}
with
\begin{equation}
q_{lm}(i) = \frac{1}{N_b} \sum_{j=1}^{N_b} Y_{lm}(\theta(\vec{r}_{ij}), \phi(\vec{r}_{ij})),
\end{equation}

\noindent where $Y_{lm}$ are spherical harmonics and $\vec{r}_{ij}$ is the distance vector between particles $i$ and $j$. We computed $q_4$ and $q_6$ using the Freud Python library~\cite{freud2020}, with Voronoi diagrams obtained via Voro++~\cite{voro++,Hernandes22}.

The CSP is defined as~\cite{PhysRevB.58.11085,Stukowski_2012}:

\begin{equation}
\mathrm{CSP} = \sum_{i=1}^{N/2} |\vec{r}_i + \vec{r}_{i+N/2}|^2,
\end{equation}

\noindent where $\vec{r}_i$ and $\vec{r}_{i+N/2}$ are vectors from the central atom to opposite neighbors, assuming $N=8$ neighbors as in a BCC structure. CSP values near zero indicate crystalline order, while broader distributions are typical of amorphous or fluid phases. When needed, we used the Polyhedral Template Matching (PTM) method as implemented in OVITO~\cite{Larsen2016,ovito} to validate our structural classification.

For simplicity, all quantities in our Results Section are reported in standard Lennard-Jones (LJ) reduced units~\cite{AllenBook}.

\section{Results and discussion}\label{results}

\begin{figure}[h!]
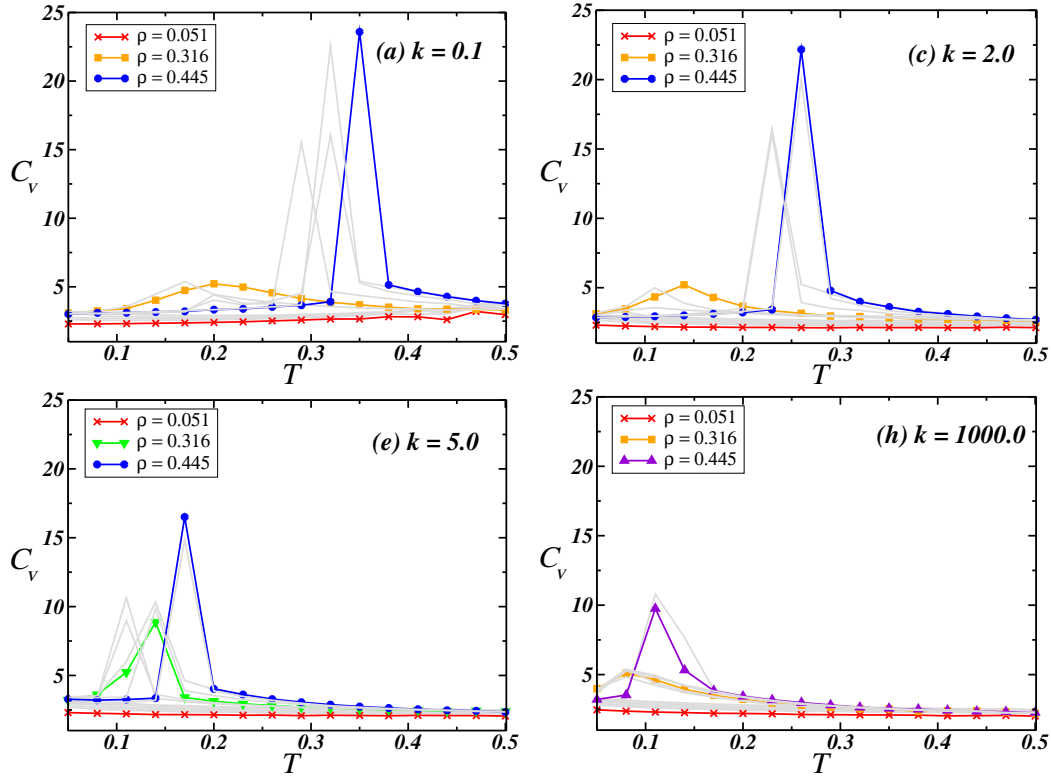

    \centering
    \includegraphics[width=0.44\textwidth]{figures/Cv_k0.1.eps} 
    \includegraphics[width=0.44\textwidth]{figures/Cv_k2.eps} 
    \includegraphics[width=0.44\textwidth]{figures/Cv_k5.eps} 
    \includegraphics[width=0.44\textwidth]{figures/Cv_k1000.eps} 
\caption{Specific heat at constant volume, $C_V$, as a function of temperature for dimeric systems with selected spring constants: (a) $k = 0.1$, (b) $k = 2.0$, (c) $k = 5.0$, and (d) $k = 1000.0$. These values represent the flexible ($k = 0.1$, $2.0$), intermediate ($k = 5.0$), and rigid ($k = 1000.0$) regimes discussed in the text. For each case, we show representative densities corresponding to different structural regimes: high ($\rho = 0.445$), intermediate ($\rho = 0.316$), and low ($\rho = 0.051$).}
    
    \label{Cv_k}
\end{figure}

We begin our analysis by examining the specific heat at constant volume, $C_V$, for selected values of the spring constant $k$ that illustrate the distinct regimes observed in our study: flexible dimers ($k = 0.1$ and $k = 2.0$), an intermediate stiffness case ($k = 5.0$), and the rigid limit ($k = 1000.0$). These values were chosen to highlight the main structural transitions and thermodynamic signatures discussed in the phase diagrams and structural analyses. The behavior of $C_V$ as a function of temperature for these representative cases is shown in Figure~\ref{Cv_k}.

First-order phase transitions are evidenced by discontinuities or sharp jumps in $C_V$, as clearly observed at the density $\rho = 0.445$ for all values of $k$. In contrast, second-order or continuous transitions 
appear as broad peaks in $C_V$, such as those identified around $\rho = 0.316$. At lower densities, for instance $\rho = 0.051$, no clear signatures of phase transitions were detected.

These results reveal a strong dependence of the solidification process on the flexibility of the dimers. Systems composed of more flexible dimers (i.e., lower $k$) undergo solid-fluid transitions at higher temperatures. As the rigidity increases, the onset of solidification shifts to lower temperatures, indicating that the mechanical coupling between monomers plays a crucial role in the thermodynamic stability of both fluid and solid phases.

Our goal is to determine the temperatures at which first- and second-order phase transitions occur, thereby distinguishing solid from fluid regions in the phase diagrams. Beyond identifying the transitions, we aim to characterize the different solid phases that emerge. To support this analysis, we employ a set of order parameters and structural metrics, using the system with spring constant $k = 2.0$ as a representative case. We focus on three characteristic densities ($\rho = 0.445$, $\rho = 0.316$, and $\rho = 0.216$)
each exhibiting distinct thermodynamic and structural behavior.

\begin{figure}[htp!]
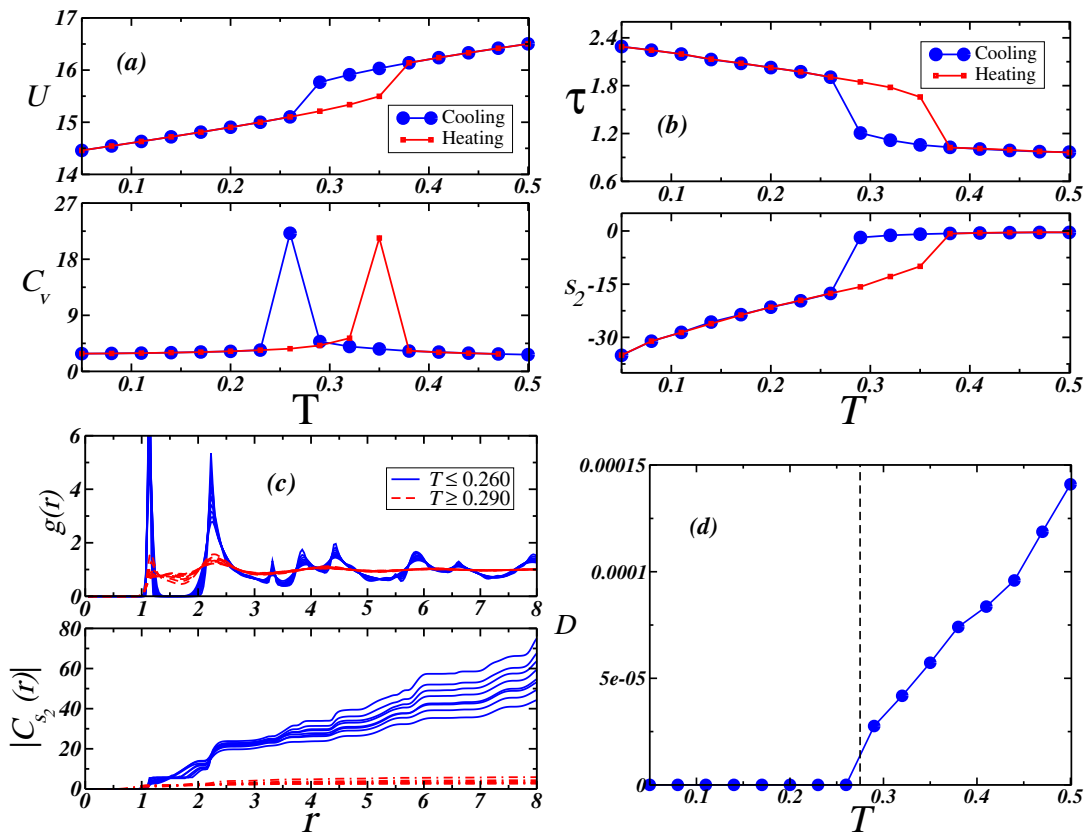

    \centering
    \includegraphics[width=0.45\textwidth]{figures/U_T_k2_L8.25_3.eps} 
    \includegraphics[width=0.45\textwidth]{figures/tau_T_k2_L8.25_3.eps} 
    \includegraphics[width=0.45\textwidth]{figures/gr_k2_L8.25_3.eps} 
    \includegraphics[width=0.45\textwidth]{figures/D_T_k_2.0_L_8.25.eps} 
    \caption{Thermodynamic and structural analysis for the system with spring constant $k = 2.0$ at density $\rho = 0.445$.
(a) Potential energy $U$ and specific heat at constant volume $C_V$ as functions of temperature $T$ during both cooling and heating processes.
(b) Translational order parameter $\tau$ and two-body excess entropy $s_2$ as functions of $T$ for both thermal paths.
(c) Radial distribution function $g(r)$ and the cumulative two-body entropy $|C_{s_2}(r)|$ as functions of interparticle distance, computed during the cooling process.
(d) Diffusion coefficient $D$ as a function of $T$ under cooling conditions.
}
    \label{U_Cv_k2_1}
\end{figure}

We begin our analysis with the highest density, $\rho = 0.445$, whose properties are shown in Fig.~\ref{U_Cv_k2_1}. Panel (a) displays the potential energy $U$ as a function of temperature $T$, 
revealing a clear discontinuity accompanied by a hysteresis loop between the cooling and heating processes - a characteristic signature of a first-order phase transition.  The corresponding
specific heat at constant volume, $C_V$, shown in the same panel,
further supports this conclusion: a sharp jump in $C_V$ occurs near $T = 0.260$ during cooling and around $T = 0.350$ during heating.
We define the transition temperature as the average between the values obtained in the cooling and heating branches.

Panel (b) shows the temperature dependence of the translational order parameter $\tau$ and the two-body entropy $s_2$, both of which also exhibit hysteresis. The translational order parameter $\tau$, which measures the degree of structural organization in the system, increases in the solid phase, decreases in the liquid phase, and tends to zero in the ideal gas limit. A sharp transition from a highly structured (solid) phase to a disordered (fluid) phase is evident near the same temperature range identified in $U$ and $C_V$. 
Similarly, the two-body entropy $s_2$, which also reflects the degree of structural order, shows a similar temperature dependence, reinforcing the occurrence of a solid-fluid transition.

To further characterize the phase change, panel (c) displays
the radial distribution function $g(r)$ and the cumulative two-body entropy $|C_{s_2}(r)|$ as functions of interparticle distance, both computed for the center of mass of the dimers during the cooling process. At higher temperatures ($T \geq 0.290$, red dashed lines), $g(r)$ exhibits typical fluid-like behavior, characterized by short-range correlations, and $|C_{s_2}(r)|$ rapidly saturating cumulative entropy. In contrast, at lower temperatures ($T \leq 0.260$, blue solid lines), $g(r)$ shows pronounced peaks in the first and second coordination shells—hallmarks of a solid phase—while $|C_{s_2}(r)|$ diverges at long distances, as expected for ordered systems. These results provide additional confirmation of a first-order phase transition.

To complement the structural analysis, panel (d) displays the diffusion coefficient $D$ of the dimer center of mass as a function of temperature. The system exhibits significant diffusive behavior only for $T \geq 0.290$, while diffusion is strongly suppressed for $T \leq 0.260$, consistent with the formation of a solid phase at low temperatures.
Together, these results provide compelling evidence for a 
first-order solid–fluid transition at this density, governed by the temperature-dependent structural organization of the system.

\begin{figure}[h!]
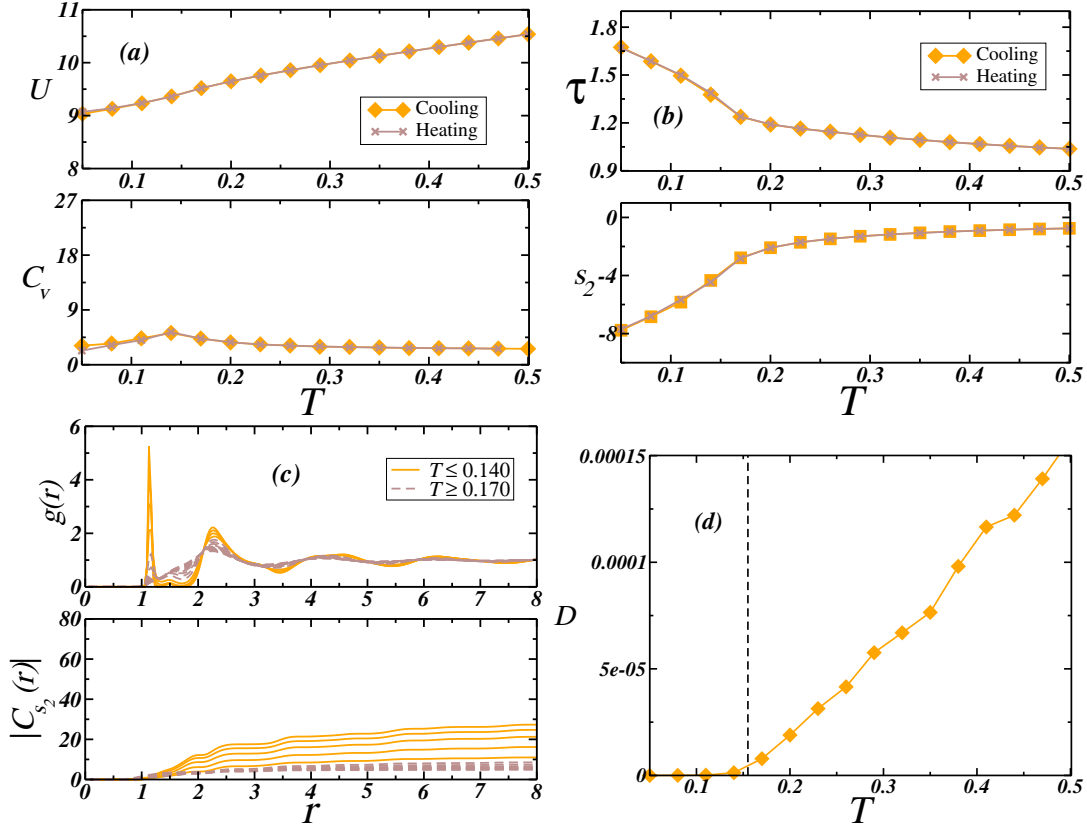

    \centering
    \includegraphics[width=0.45\textwidth]{figures/U_T_k2_L9.25_3.eps} 
    \includegraphics[width=0.45\textwidth]{figures/tau_T_k2_L9.25_3.eps} 
    \includegraphics[width=0.45\textwidth]{figures/gr_k2_L9.25_3.eps} 
    \includegraphics[width=0.45\textwidth]{figures/D_T_k_2.0_L_9.25.eps} 
    \caption{Thermodynamic and structural analysis for the system with spring constant $k = 2.0$ at density $\rho = 0.316$.
(a) Potential energy $U$ and specific heat at constant volume $C_V$ as functions of temperature $T$ during cooling and heating cycles.
(b) Translational order parameter $\tau$ and two-body excess entropy $s_2$ as functions of $T$ for both thermal protocols.
(c) Radial distribution function $g(r)$ and cumulative two-body entropy $|C_{s_2}(r)|$ as functions of interparticle distance, obtained during the cooling process.
(d) Diffusion coefficient $D$ as a function of $T$ under cooling conditions.}
    \label{U_Cv_k2_2}
\end{figure}

Intermediate densities exhibit a different type of phase transition, as illustrated in Fig.~\ref{U_Cv_k2_2} for $\rho = 0.316$. Panel (a) shows the potential energy $U$ as a function of temperature $T$. In contrast to the higher-density case, no discontinuities are observed,
and the energy profiles for both cooling and heating processes coincide, indicating the absence of hysteresis. However, the specific heat at constant volume, $C_V$, displays a well-defined peak at $T = 0.140$, rather than a discontinuity, indicating the occurrence of a second-order (continuous) phase transition.

Structural indicators further support this interpretation. Panel (b) presents the translational order parameter $\tau$ and the two-body entropy $s_2$, both exhibiting smooth, continuous behavior across the transition. While their values differ between
the low temperature ($T \leq 0.140$) and high-temperature
($T \geq 0.170$) regimes, no abrupt changes or signs of hysteresis are observed, consistent with a continuous transformation.

Panel (c) displays the radial distribution function $g(r)$ and the cumulative two-body entropy $|C_{s_2}(r)|$ as functions of interparticle distance, computed for the center of mass of the dimers during the cooling process. For $T \leq 0.140$, $g(r)$ exhibits a  prominent nearest-neighbor peak around $r \approx 1.2$, followed by a rapid decay and small oscillations about unity—hallmarks of an amorphous solid, which lacks long-range positional order. At higher temperatures ($T \geq 0.170$), $g(r)$ assumes the typical profile of a fluid, with broader, less pronounced peaks. The gradual change in $|C_{s_2}(r)|$ with temperature further supports the continuous nature of the structural transformation.

Finally, panel (d) presents the diffusion coefficient $D$ of the dimer center of mass as a function of temperature. While the system exhibits diffusive behavior for $T \geq 0.170$, while diffusion is significantly reduced below $T = 0.140$, reinforcing the presence of a solid phase at low temperatures. However, the absence of crystalline order, as suggested by the structural functions, indicates that this low-temperature phase corresponds to an amorphous solid.

\begin{figure}[h!]
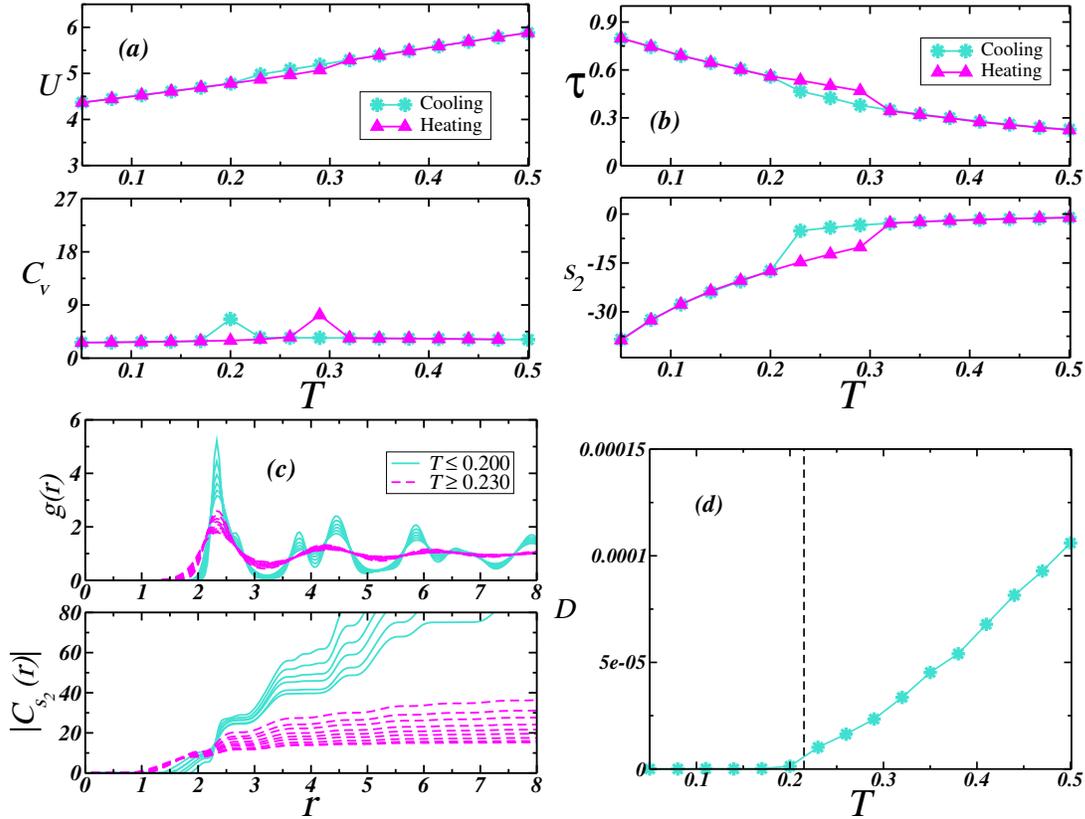

    \centering
    \includegraphics[width=0.45\textwidth]{figures/U_T_k2_L10.5_3.eps} 
    \includegraphics[width=0.45\textwidth]{figures/tau_T_k2_L10.5_3.eps} 
    \includegraphics[width=0.45\textwidth]{figures/gr_k2_L10.5_3.eps} 
    \includegraphics[width=0.45\textwidth]{figures/D_T_k_2.0_L_10.5.eps} 
    \caption{Thermodynamic and structural analysis for the system with spring constant $k = 2.0$ at density $\rho = 0.216$.
(a) Potential energy $U$ and specific heat at constant volume $C_V$ as functions of temperature $T$ during cooling and heating cycles.
(b) Translational order parameter $\tau$ and two-body excess entropy $s_2$ as functions of $T$ for both thermal protocols.
(c) Radial distribution function $g(r)$ and cumulative two-body entropy $|C_{s_2}(r)|$ as functions of interparticle distance, obtained during the cooling process.
(d) Diffusion coefficient $D$ as a function of $T$ under cooling conditions.}
    \label{U_Cv_k2_3}
\end{figure}

The next analysis, analogous to the previous cases, was performed for $\rho = 0.216$, as shown in Figure~\ref{U_Cv_k2_3}. Similar to the case of $\rho = 0.445$, this system also exhibits a first-order phase transition. Panel (a) shows the potential energy $U$ as a function of temperature $T$, where a discontinuity followed by a hysteresis loop between the cooling and heating curves is observed—although less pronounced than in the higher-density case. The specific heat at constant volume $C_V$, presented in the same panel, confirms this behavior, displaying discontinuities around $T = 0.200$ during cooling and $T = 0.230$ during heating. As in previous cases, the transition temperature is defined as the average of these two values.

Panel (b) displays the translational order parameter $\tau$ and the two-body entropy $s_2$ as functions of temperature. Both parameters exhibit hysteresis and clearly indicate a transition from an ordered (solid) to a disordered (fluid) state, consistent with a first-order transition.

Further evidence comes from panel (c), which shows the radial distribution function $g(r)$ and the cumulative two-body entropy $|C_{s_2}(r)|$ for the dimer center of mass during the cooling process. The $g(r)$ profile is consistent with a solid-like phase at low temperatures; however, the structural organization differs from that observed at $\rho = 0.445$. While the first coordination shell dominates the structure at $\rho = 0.445$ ($r \approx 1.2$), for $\rho = 0.216$ the first prominent peak appears near $r \approx 2.0$, suggesting the formation of a different solid phase. In agreement, $|C_{s_2}(r)|$ diverges for $T \leq 0.200$, as expected for ordered structures, and converges for $T \geq 0.230$, characteristic of fluid behavior.

Lastly, panel (d) presents the diffusion coefficient $D$ as a function of temperature. The system displays diffusive behavior only for $T \geq 0.230$, while diffusion is suppressed below $T = 0.200$, further supporting the identification of a solid phase at low temperatures.

The analyses described above were repeated for all densities and for all values of the spring constant $k$, in order to determine the temperatures at which first- and second-order phase transitions occur. In cases where hysteresis was present, the transition temperature was defined as the average between the temperatures observed during cooling and heating. In the absence of hysteresis, the transition temperature was taken as the temperature corresponding to the peak (for second-order) or discontinuity (for first-order) in the specific heat $C_V$. These evaluations were systematically carried out along all isochores.

\begin{figure}[h!]
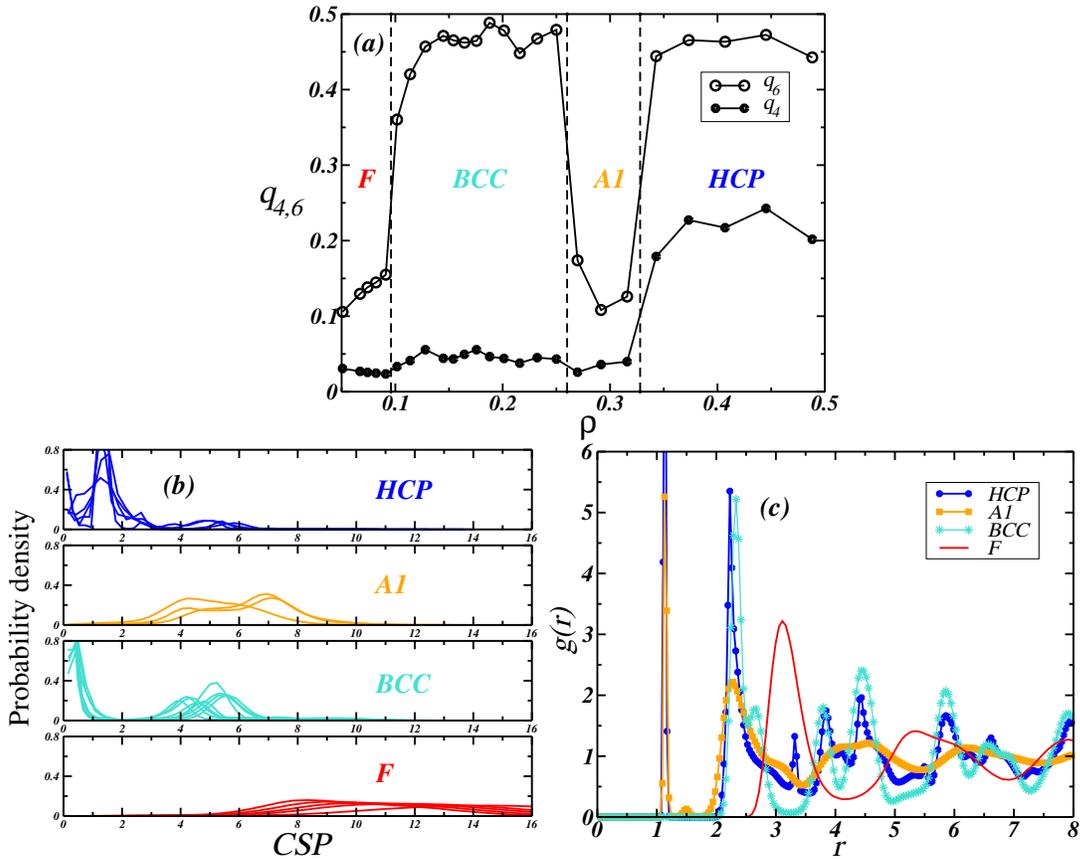

    \centering
    \includegraphics[width=0.5\textwidth]{figures/q6_q4_k2_T0.050.eps} \\
    \includegraphics[width=0.45\textwidth]{figures/csp_k2_T0.050_2.eps} 
    \includegraphics[width=0.45\textwidth]{figures/gr_k2_T0.050.eps} 
    \caption{Structural characterization of the system with spring constant $k = 2.0$ at fixed temperature $T = 0.050$.
(a) Orientational order parameters $q_4$ and $q_6$ as functions of density $\rho$.
(b) Probability density distribution of the centrosymmetry parameter (CSP), calculated considering the eight nearest neighbors.
(c) Radial distribution function $g(r)$ computed from the center of mass of the dimers for selected densities.}
    \label{q4q6_csp_1}
\end{figure}

We now turn to the analysis of isotherms in order to explore additional structural parameters that can help classify the distinct solid phases previously identified. To this end, we employ the orientational order parameter (OOP), the centrosymmetry parameter (CSP), and the radial distribution function $g(r)$ to characterize and distinguish between the different solid structures observed along the isotherms.

We begin this structural analysis with the system defined by a spring constant of $k = 2.0$. Figure~\ref{q4q6_csp_1}, panel (a), shows the orientational order parameters $q_4$ and $q_6$ as functions of density at a fixed temperature of $T = 0.050$. Based on the behavior of these parameters, four distinct regimes can be identified. At high densities, $q_4 \approx 0.2$ and $q_6 \approx 0.5$, values consistent with a well-ordered hexagonal close-packed (HCP) structure. As the density decreases, both $q_4$ and $q_6$ decline, indicating a loss of orientational order and confirming the presence of the amorphous solid phase previously observed at $\rho = 0.316$ (see Figure~\ref{U_Cv_k2_2}). Upon further density reduction, $q_4$ remains low while $q_6$ increases again, suggesting the emergence of a distinct crystalline structure, identified here as a body-centered cubic (BCC) phase. Finally, at the lowest densities, both $q_4$ and $q_6$ drop significantly, signaling a transition to a disordered fluid phase.

Panel (b) presents the probability density of the centrosymmetry parameter (CSP), computed using the eight nearest neighbors, as appropriate for identifying BCC structures. In a perfect BCC crystal, CSP values are close to zero. Other crystalline phases also yield low CSP values, while amorphous and fluid phases typically exhibit broader, more dispersed distributions. The uppermost profile (blue) shows a sharp distribution centered around CSP $\approx 1.0$, indicative of a highly ordered HCP phase. The following distributions (orange) are broad and unstructured, consistent with an amorphous solid phase (labeled A1). The cyan curve displays a dominant peak near zero but also significant weight around CSP values of 4.0 to 5.0, reflecting a BCC-like phase with noticeable structural imperfections. Finally, the lower profiles (red) are widely spread, characteristic of disordered fluid phases (F).

Panel (c) presents the radial distribution function $g(r)$ for representative densities. At $\rho = 0.445$, the HCP structure exhibits well-defined peaks corresponding to the first and second coordination shells, indicative of long-range crystalline order. The amorphous solid (A1) at $\rho = 0.316$ shows a dominant first peak followed by smooth, damped oscillations—typical of disordered, non-crystalline solids. At $\rho = 0.216$, the BCC-like phase with imperfections shows a distinct peak at the second coordination shell ($r \approx 2.0$), in contrast to the HCP structure, and exhibits additional peaks at larger distances, supporting the presence of a different crystalline symmetry. Finally, the fluid phase (F) at $\rho = 0.091$ displays only short-range order, with a rapidly decaying $g(r)$, as expected for disordered liquids.

Unlike the isochoric analysis, as exemplified by the $k = 2.0$ case, which primarily focused on identifying the nature (first- or second-order) of phase transitions, the isothermal analysis provides deeper insights into the structural characteristics of the various phases. This approach not only detects phase transitions but also enables the classification of the resulting solid phases, revealing more nuanced structural behavior. To further illustrate this, we now examine two representative systems with spring constants $k = 5.0$ and $k = 100.0$, which exhibit distinct solid-phase symmetries.

\begin{figure}[h!]
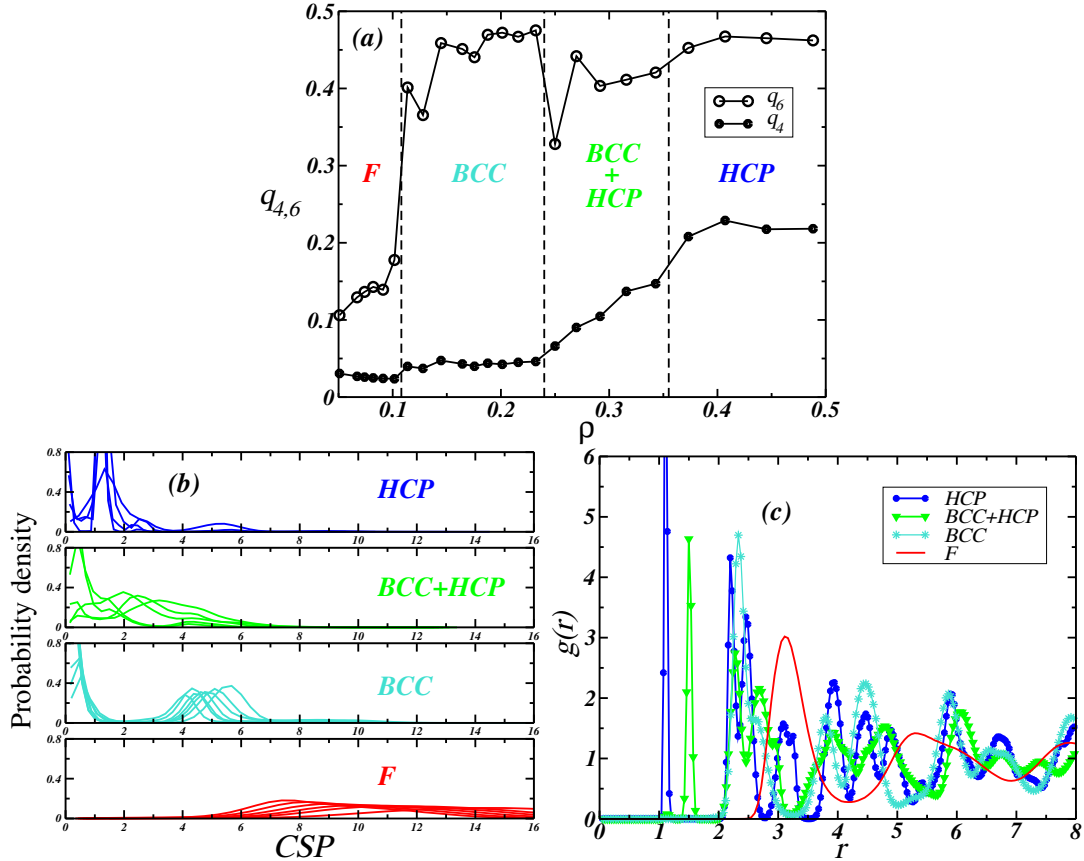

    \centering
    \includegraphics[width=0.5\textwidth]{figures/q6_q4_k5_T0.050.eps} \\
    \includegraphics[width=0.45\textwidth]{figures/csp_k5_T0.050_2.eps} 
    \includegraphics[width=0.45\textwidth]{figures/gr_k5_T0.050.eps} 
    \caption{Structural characterization of the system with spring constant $k = 5.0$ at fixed temperature $T = 0.050$.
(a) Orientational order parameters $q_4$ and $q_6$ as functions of density $\rho$.
(b) Probability density distribution of the centrosymmetry parameter (CSP), calculated using the eight nearest neighbors.
(c) Radial distribution function $g(r)$ calculated from the center of mass of the dimers for selected densities.}
    \label{q4q6_csp_2}
\end{figure}

We now examine the system with spring constant $k = 5.0$ at a fixed temperature of $T = 0.050$, as shown in Figure~\ref{q4q6_csp_2}. The orientational order parameters $q_4$ and $q_6$, presented in panel (a), exhibit noticeable differences compared to the $k = 2.0$ case. Notably, no intermediate amorphous region is observed. Instead, the data suggest the presence of a coexistence region between hexagonal close-packed (HCP) and body-centered cubic (BCC) solid phases. While the HCP, BCC, and fluid (F) phases exhibit behaviors consistent with those previously identified, the intermediate region presents a distinct profile: the $q_4$ values are significantly higher than those typically associated with disordered structures, indicating that this regime does not correspond to an amorphous phase.

This interpretation is corroborated by the centrosymmetry parameter (CSP), shown in panel (b). The CSP distributions for the HCP (blue), BCC (cyan), and fluid (red) phases are qualitatively similar to those observed for the $k = 2.0$ system. However, the intermediate distribution (green) lies between the profiles for HCP and BCC, consistent with a structural coexistence or a mixed-phase region involving both crystalline symmetries.

A similar trend is evident in the radial distribution functions presented in panel (c). The HCP, BCC, and fluid phases display the expected $g(r)$ profiles, reflecting their respective degrees of structural order and symmetry. The intermediate region, however, shows features that are intermediate between the HCP and BCC patterns, further supporting the interpretation of phase coexistence rather than amorphization.

\begin{figure}[h!]
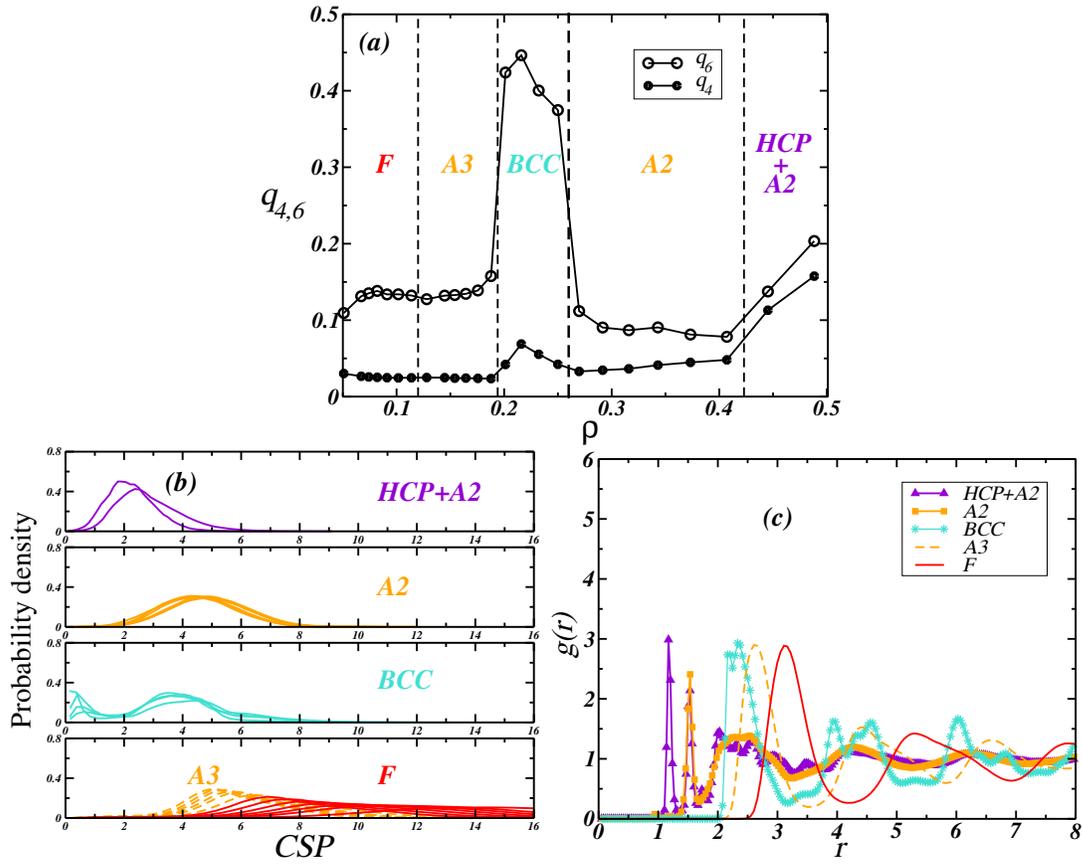

    \centering
    \includegraphics[width=0.5\textwidth]{figures/q6_q4_k100_T0.050.eps} \\
    \includegraphics[width=0.45\textwidth]{figures/csp_k100_T0.050_2.eps} 
    \includegraphics[width=0.45\textwidth]{figures/gr_k100_T0.050.eps} 
    \caption{Structural characterization of the system with spring constant $k = 100.0$ at fixed temperature $T = 0.050$.
(a) Orientational order parameters $q_4$ and $q_6$ as functions of density $\rho$.
(b) Probability density distribution of the centrosymmetry parameter (CSP), computed using the eight nearest neighbors.
(c) Radial distribution function $g(r)$ of the dimer center of mass for selected densities.}
    \label{q4q6_csp_3}
\end{figure}

We now present the results for the system with spring constant $k = 100.0$, as shown in Figure~\ref{q4q6_csp_3}. The analysis of the orientational order parameters $q_4$ and $q_6$ reveals, at high densities, a coexistence between a hexagonal close-packed (HCP) structure and an amorphous solid phase, here referred to as A2. This is followed by an extended density range dominated by the A2 amorphous phase. At intermediate densities, a relatively narrow region is identified where a body-centered cubic (BCC) structure emerges. Below this BCC region, another short density range appears, characterized by a distinct amorphous solid phase (A3), which is associated with a peak in the $C_V$ curve. Finally, at low densities, the system transitions into a disordered fluid (F) phase.

Panel (b) shows the probability density distributions of the centrosymmetry parameter (CSP). At high densities, the CSP profile resembles that of an amorphous structure but with lower overall values than typical for fully disordered phases, suggesting a partially disordered structure with localized HCP-like ordering. The remaining regions—corresponding to the A2 and A3 amorphous phases, the BCC phase, and the fluid phase—exhibit CSP distributions consistent with their respective structural assignments.

The radial distribution functions $g(r)$, shown in panel (c), further support these interpretations. The HCP+A2 coexistence region exhibits a $g(r)$ profile that is predominantly amorphous in character. Importantly, the A2 and A3 amorphous phases display distinct features, particularly in the position and shape of the first peak in $g(r)$, indicating clear structural differences between them. Meanwhile, the BCC and fluid phases present $g(r)$ profiles consistent with well-defined crystalline and disordered liquid structures, respectively.

\begin{figure}[h!]
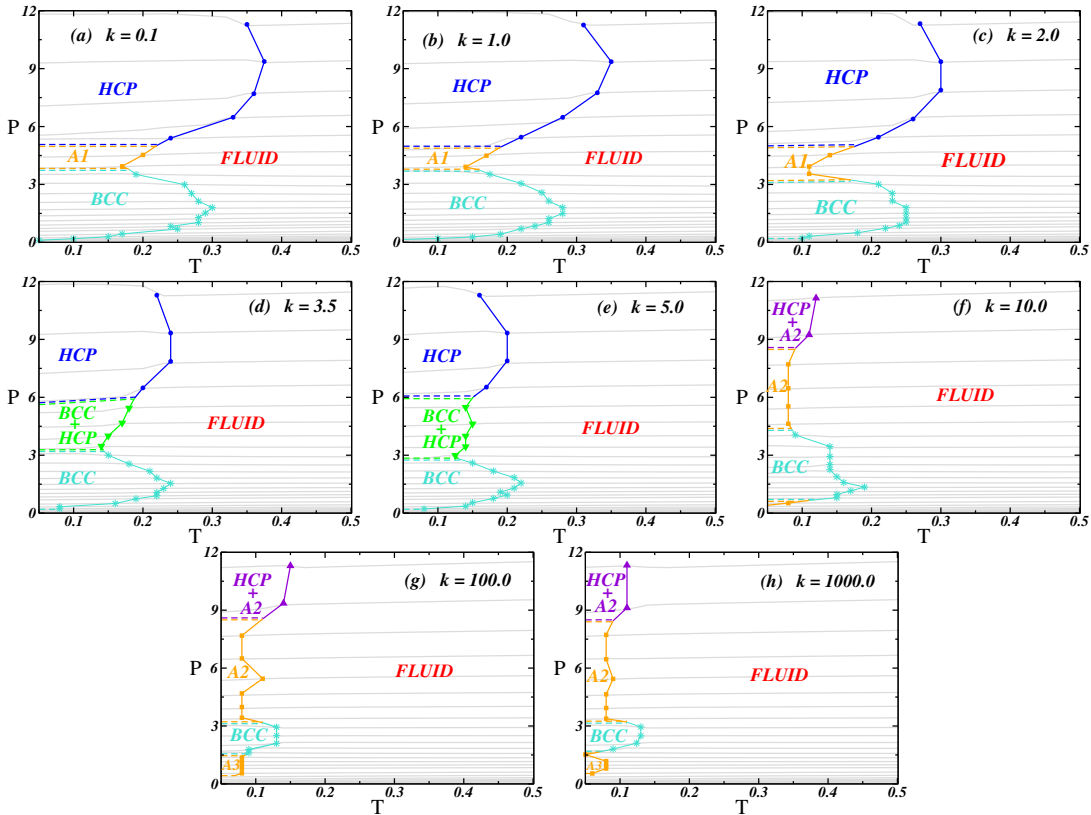

    \centering
    \includegraphics[width=0.3\textwidth]{figures/diagram_k0.1.eps} 
    \includegraphics[width=0.3\textwidth]{figures/diagram_k1.eps}
    \includegraphics[width=0.3\textwidth]{figures/diagram_k2.eps} 
    \includegraphics[width=0.3\textwidth]{figures/diagram_k3.5.eps} 
    \includegraphics[width=0.3\textwidth]{figures/diagram_k5.eps} 
    \includegraphics[width=0.3\textwidth]{figures/diagram_k10.eps}
    \includegraphics[width=0.3\textwidth]{figures/diagram_k100.eps} 
    \includegraphics[width=0.3\textwidth]{figures/diagram_k1000.eps} 
    \caption{Pressure–temperature ($P$–$T$) phase diagrams for systems with different spring constants.
Panels (a)–(h) correspond to spring constants of $k = 0.1$, $1.0$, $2.0$, $3.5$, $5.0$, $10.0$, $100.0$, and $1000.0$, respectively. Gray lines represent the simulated isochores, while phases are identified based on discontinuities or anomalies in thermodynamic and structural observables.}
    \label{PT_diagrams}
\end{figure}

By analyzing all the systems studied, we identify three distinct structural profiles that emerge as a function of dimer rigidity. The first group comprises systems with low spring constants, namely $k = 0.1$, $1.0$, and $2.0$, corresponding to highly flexible dimers. These systems exhibit phase diagrams that closely resemble those of monomeric particles~\cite{Oliveira06a,Oliveira06b,Bordin2023a}, displaying an HCP solid phase at high densities, followed by a narrow region of an amorphous solid phase (A1), a broad BCC solid region, and finally a fluid phase, as the density decreases at fixed temperature.

The second group includes systems with intermediate spring constants, such as $k = 3.5$ and $5.0$. These systems deviate slightly from the flexible regime by exhibiting a region of structural coexistence between HCP and BCC phases, replacing the distinct amorphous region observed in the first group.

The third group corresponds to rigid dimers, represented by systems with spring constants $k = 10.0$, $100.0$, and $1000.0$. These systems exhibit more complex structural behavior: a coexistence region between an HCP structure and a second amorphous phase (A2) at high densities, followed by an extended A2 region, a narrow BCC solid phase, a third amorphous phase (A3), and ultimately a transition to the fluid phase. Although the system with $k = 10.0$ is not fully rigid, its structural behavior aligns with the trend observed for the stiffer dimers.

These findings are summarized in Figure~\ref{PT_diagrams}, which presents the pressure–temperature ($P$–$T$) phase diagrams for systems with spring constants: (a) $k = 0.1$, (b) $k = 1.0$, (c) $k = 2.0$, (d) $k = 3.5$, (e) $k = 5.0$, (f) $k = 10.0$, (g) $k = 100.0$, and (h) $k = 1000.0$.

\begin{figure}[h!]
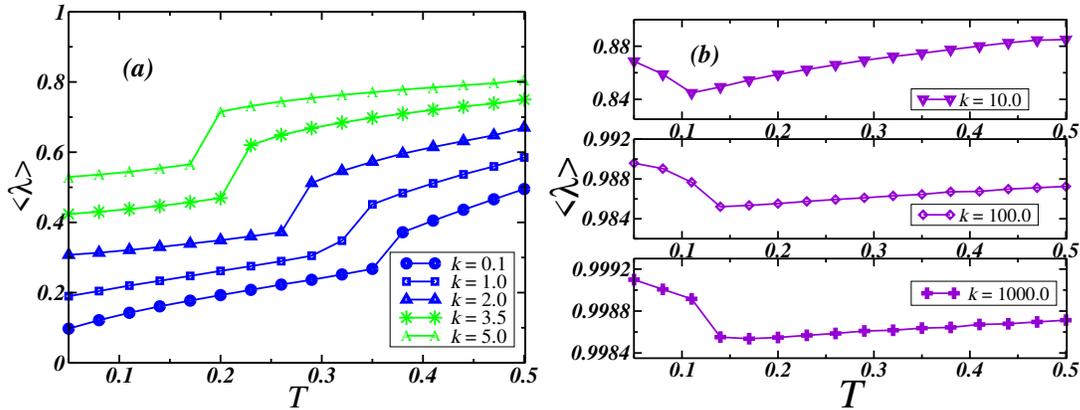

    \centering
    \includegraphics[width=0.45\textwidth]{figures/sep_k_weak.eps} 
    \includegraphics[width=0.45\textwidth]{figures/sep_k_strong.eps} 
    \caption{Average monomer-monomer separation $\langle\lambda\rangle$, computed as an average over all dimers in the system, as function of temperature $T$ for a fixed density of $\rho = 0.445$. In (a), we show flexible dimers ($k = 0.1$, $1.0$ and $2.0)$ and intermediate dimers ($k = 3.5$ and $5.0$) and in (b) we show stiffer dimers ($k = 10.0$, $100.0$, and $1000.0$). Error bars are smaller than the data points.}
    \label{separation}
\end{figure}

To understand why rigid dimers exhibit a broader range of amorphous phases and a coexistence between amorphous and HCP solids—rather than forming a single, well-defined HCP structure—we analyze the average monomer–monomer separation, $\langle \lambda \rangle$, calculated as an average over all dimers in the system. Figure~\ref{separation} presents $\langle \lambda \rangle$ as a function of temperature $T$ at fixed density $\rho = 0.445$. Panel (a) shows results for flexible ($k = 0.1$, $1.0$, and $2.0$) and intermediate-stiffness dimers ($k = 3.5$ and $5.0$), while panel (b) displays the behavior for stiffer dimers ($k = 10.0$, $100.0$, and $1000.0$). In all cases, a first-order phase transition is observed—consistent with our earlier analyses at high densities—though occurring at different temperatures depending on the spring constant $k$.

The most striking observation lies in the contrasting trends of $\langle \lambda \rangle$ across the phase transition. For flexible and intermediate-stiffness dimers, the monomer–monomer separation is smaller in the solid phase and increases upon melting, indicating that flexibility facilitates tighter packing in the crystalline state. In contrast, for stiffer dimers, this behavior is inverted: $\langle \lambda \rangle$ is larger in the solid phase than in the fluid, suggesting that rigidity hinders efficient packing and induces geometric frustration. This frustration stems from the inability of rigid dimers (with large $k$) to adjust their bond lengths to accommodate local packing constraints. While flexible dimers can compress or extend to conform to their surroundings—thus favoring crystalline order—rigid dimers maintain nearly fixed internal separations, which may be incompatible with optimal packing, especially under high-density conditions. This mechanism helps explain the broader stability of the HCP phase in flexible dimers and the dominance of amorphous structures or HCP–amorphous coexistence in rigid ones.

At intermediate densities, however, the behavior becomes less regular. Although the inversion in $\langle \lambda \rangle$ still appears in several cases, it is not universal. Its presence depends on a complex interplay between dimer rigidity, local packing efficiency, and thermal fluctuations. In particular, in amorphous solids—where local environments vary significantly and long-range order is absent—the impact of rigidity may be either enhanced or suppressed. These findings indicate that geometric frustration induced by dimer stiffness plays a central role in determining the structural organization at high densities, but its influence is modulated by both structural and dynamic factors as the density decreases.

\section{Conclusion}\label{conclusions}

In this work, we investigated the impact of bond flexibility on
the phase behavior of dimers interacting via a two-length-scale core-softened potential. Through molecular dynamics simulations involving both heating and cooling cycles, we identified both first-order solid–fluid transitions marked by energy hysteresis and discontinuities in the specific heat at constant volume. We also observed second-order transitions between amorphous solids and fluids, characterized by smooth changes in thermodynamic quantities and peaks in the specific heat. A detailed structural analysis using orientational and local order parameters enabled us to distinguish between hexagonal close-packed (HCP) and body-centered cubic (BCC) crystalline phases, amorphous solids, and mixed-phase regions including HCP–BCC and HCP–amorphous coexistence.

Our findings reveal that dimer flexibility plays a central role in determining structural order. Highly flexible dimers are able to adjust their internal separation, which favors efficient packing and promotes the formation of extended crystalline phases. As a result, these systems display broad regions of HCP ordering at high densities, a narrow amorphous zone at intermediate densities, and a wide BCC phase at lower densities—closely resembling the phase behavior observed in monomeric core-softened fluids~\cite{Bordin2023a}.

In contrast, stiff dimers, which cannot deform to accommodate local packing constraints, show a qualitatively different scenario. Their phase diagrams are dominated by geometric frustration: instead of a fully ordered HCP structure at high densities, we observe a coexistence between amorphous and HCP-like arrangements. Additionally, these systems exhibit reentrant amorphous phases at both low and intermediate densities. The inability of rigid dimers to optimize local packing leads to the suppression of long-range crystalline order and favors the formation of structurally disordered, glasslike phases. Overall, our results highlight internal flexibility as a key design parameter for tuning structural order, crystallization propensity, and the emergence of amorphous behavior in soft-matter systems governed by core-softened interactions.

Finally, several extensions and improvements to the current dimer model and methodology 
merit future exploration. One promising direction involves studying the effect of confinement on
dimers interacting via core-softened potentials, particularly in the presence of structured environments such as grafted polymer brushes. 
Additionally, investigating the phase behavior
of mixtures comprising dimers and trimers could yield further insights into the role of 
molecular architecture in complex fluids. 
Research along these lines is currently underway in our laboratory and will be reported in 
future publications.

\section*{Acknowledgments}

J.R.B. acknowledges financial support from CNPq (grant numbers 405479/2023-9, 441728/2023-5, and 304958/2022-0A). DFKS acknowledges support from CAPES (Finance Code 001). V.M.T. thanks the financial support provided by SECIHTI-Mexico through the program "Convocatoria Ciencia Básica y de Frontera 2023-2024," grant number CBF2023-2024-2725.

\section*{References}
\bibliographystyle{iopart-num}
\bibliography{biblio}

\end{document}